# Ultrasonic spectroscopy of sessile droplets coupled to optomechanical sensors

K. G. Scheuer[a], F. B. Romero[b], G. J. Hornig[b], R. G. DeCorby[b]*

[a]Ultracoustics Technologies Ltd., Sherwood Park, AB, Canada, T8A 3H5
[b]ECE Department, University of Alberta, 9211-116 St. NW, Edmonton, AB, Canada, T6G 1H9

We describe a system for interrogating the acoustic properties of sub-nanoliter liquid samples within an open microfluidics platform. Sessile droplets were deposited onto integrated optomechanical sensors, which possess ambient-medium-noise-limited sensitivity and can thus passively sense the thermally driven acoustic spectrum of the droplets. The droplet acoustic breathing modes manifest as resonant features in the thermomechanical noise spectrum of the sensor, in some cases hybridized with the sensor's own vibrational modes. Excellent agreement is found between experimental observations and theoretical predictions, over the entire ~ 0 – 40 MHz operating range of our sensors. With suitable control over droplet size and morphology, this technique has the potential for precision acoustic sensing of small-volume biological and chemical samples.

## 1. Introduction

The study and manipulation of acoustic phonons, for both fundamental research and applied nanotechnology, is a topic of growing interest [1]. Nevertheless, the field of 'ultrasonic spectrometry', where the frequency-dependent absorption and velocity of acoustic waves is used as a 'label-free' method to characterize a fluid, has a long history in physics, chemistry, and applied sciences [2,3]. Often the sample container is designed to function simultaneously as an acoustic resonator, so that the loss and velocity information can be extracted from the positions and linewidths of the resonant frequencies. Typically, piezoelectric transducers are used, and the ultrasound signals are coupled to macroscopic (although sometimes very small [4]) sample volumes. The literature on this topic spans a wide range of fundamental and applied studies and uses a range of ultrasonic frequencies (KHz to GHz) [3].

In micro- and nano-scale systems, surface-acoustic-wave (SAW)-based sensors have been widely studied [5,6]. Similar to the bulk devices mentioned above, SAW-based sensors detect changes in the physical properties (e.g., velocity, loss, resonant frequency) of an acoustic wave related to some measurable quantity of interest (e.g., mass, temperature, or binding of a target analyte). Within microfluidic and lab-on-a-chip platforms, however, the main recent application of SAW devices is for manipulation and actuation of droplets or particles [7,8]. Meanwhile, there is also much interest in so-called 'open microfluidics platforms' [9,10], where sessile liquid droplets are used as 'containers' for chemical and biological entities. Potential advantages include compatibility with extremely small sample volumes and unique capabilities for analyte concentration (e.g., driven by evaporation of a solvent), chemical reactions and sensing within droplets. To date, the predominant sensing modalities in these systems have been based on electrical (e.g., conductivity) or optical (e.g., fluorescence) techniques. Here, we demonstrate that optomechanical sensors can be used to passively interrogate the micro-scale environment of a droplet.

We use our recently reported [11] high-frequency ultrasound sensors, which have state-of-the-art sensitivity for operation in either air or water. They are all-optical (optomechanical) devices (with laser-based readout) [12,13], and offer significant advantages over conventional electrical/piezo-electrical based ultrasound sensors. Firstly, their performance is not limited by electrical noise. In fact, they can operate (in both fluid and gas media) at fundamental limits set by the thermal Brownian motion of the ambient medium itself. Aside from implying that very weak ultrasound signals can be reliably detected [13], this has profound implications in terms of the potential for these sensors to passively monitor their environment, as evidenced by the results presented here. Secondly, the devices are extremely small (~ 100 μm) and can be manufactured by the thousands in a monolithic wafer-scale process. For comparable sensitivity, a piezoelectric sensor needs to be on the order of centimeters in size [14,15]. The small size of the sensors imparts an inherent omnidirectional property at very high ultrasound frequencies [14], and more importantly for the present work, enables them to sense acoustic signals within a micro-scale environment.

The current work builds on that in Ref. [16], where we detected the volume 'breathing' modes of low-profile glycerol droplets deposited onto our sensors. Those modes manifested as relatively weak Fano-resonance features in the thermomechanical noise spectrum of the sensor, limited by the high viscosity of the glycerol medium (i.e., which dampens the optomechanical sensor response) and the low-profile (i.e., spherical cap) shape of the droplets. In the present work, we employed less viscous liquids (e.g., ethylene glycol and water) combined with an increased hydrophobicity of the sensor surface to produce nearly hemispherical droplets, resulting in a greatly enhanced coupling of the droplet acoustic modes to the sensor. Excellent agreement between theory and experiment is demonstrated for a wide variety of droplet sizes, evincing that these systems have the potential to enable passive acoustic sensing (i.e., vibrational spectroscopy) of thermal Brownian motions in an open microfluidics platform.

## 2. Experimental Methods

Our optomechanical sensors are high-finesse ($F > 10^3$), spherical mirror (i.e., plano-concave Fabry-Perot) cavities fabricated in a monolithic process on a quartz wafer. They are formed by a controlled thin-film buckling process [17] which results in a partially evacuated region embedded between a lower planar Bragg mirror and an upper curved (i.e., buckled) mirror. The buckled mirror also functions as a flexible membrane exhibiting relatively high-Q mechanical vibrational modes (e.g., $Q \sim 100$ in air) in the MHz-frequency range, while also exhibiting extreme sensitivity to external static [18] and dynamic [19] pressure variations. Since this membrane mirror also forms one end of a high-finesse, spherical mirror optical resonator (i.e., with Gaussian beam cavity modes), changes in its position are easily and sensitively read out by locking the wavelength of an interrogation laser near a resonant line of the cavity (i.e., the so-called 'tuned-to-slope' technique [12-14,16]). This results in a direct mapping of mirror motion to the intensity of the light reflected by the cavity. An extensive description of the use of our cavities for high-frequency ultrasound sensing was reported elsewhere [11], and representative optical and mechanical properties are summarized for easy reference in the supplementary information file.

The top surface of the sensor chips is typically a relatively hydrophilic amorphous silicon layer. By depositing and patterning a hydrophobic fluorocarbon layer, we were previously able to achieve precise alignment between arrays of glycerol droplets and sensors [16]. Here, our goal was to achieve a stronger coupling between the droplet modes and the sensors. To this end, we uniformly coated sensor chips with a thin (~200 nm thick) hydrophobic layer, deposited using the fluorocarbon deposition step from a Bosch-process reactive ion etching (RIE) system. Small droplets of ethylene glycol (EG), water, or EG-water mixtures were then aligned and dispensed manually using a micro-syringe with a 30-gauge needle, attached to a three-axis micropositioner. Both EG and water formed nearly hemispherical drops (i.e., contact angles near 90 degrees) on the hydrophobic surface, although with some reduction in contact angle for the smallest droplets as described below. While analogous results were observed for water droplets, we restrict the discussion below to EG droplets, since their much lower evaporation rate was better suited to controlled and repeatable data collection and analysis.

A microscope photograph of a typical array of sensors, including one covered by a sessile droplet, is shown in Fig. 1(a). Schematic representations of the sensing platform are depicted in Figs. 1(b) and (c). As depicted in Fig. 1(c), the acoustic (i.e., compressive 'breathing') modes of a droplet are coupled to the underlying sensor. As shown below, these modes produce distinct and predictable resonance features within the thermomechanical noise spectrum associated with vibrational motion of the membrane mirror in our cavities, which is in turn read out by a low-power interrogation laser coupled through the quartz substrate (see the supplementary information file for additional details of the experimental setup).

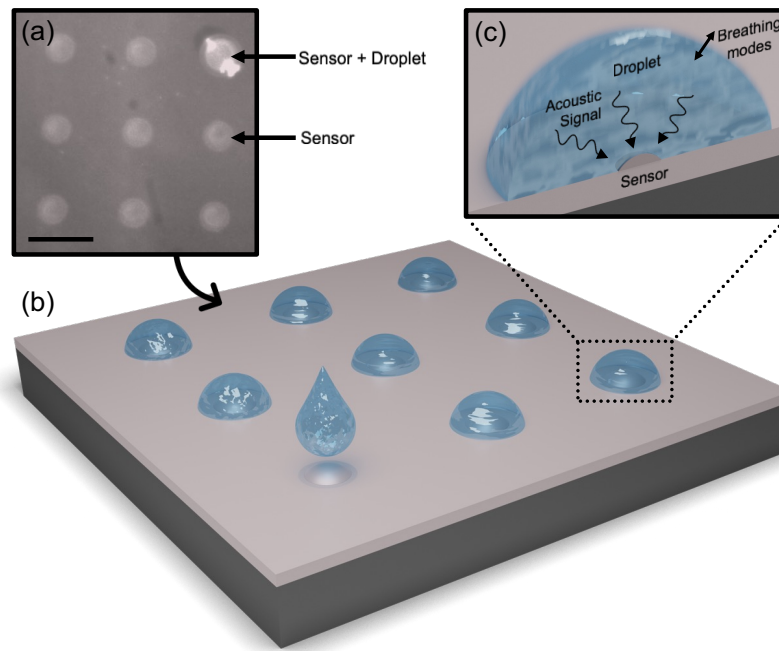

**Fig. 1.** (a) An array of 9 optomechanical sensors is shown, including one sensor (upper right) covered by a nearly hemispherical EG droplet. Scale bar: 200 µm. (b) Artistic depiction of a platform for acoustic sensing of sessile droplets using an array of optomechanical sensors. (c) Artistic depiction of a single droplet/sensor combination, illustrating the coupling of droplet acoustic (breathing) modes to the underlying sensor.

## 3. Results

We begin with results for a smaller droplet, motivated by its relatively sparse acoustic spectrum, which makes the fitting and assignment of modes more straightforward. The thermomechanical noise spectrum of a sensor lying under a small droplet is shown by the red curve in Fig. 2(a), while the noise spectrum for the same sensor covered by a large amount of EG (i.e., a 'bulk' EG medium) is shown by the blue curve. The datasets used to produce these curves were averaged and smoothed, as described in the supplementary information file. Microscope top- and side-view photographs of the associated droplet/sensor combination are shown in Figs. 2(b) and 2(c), respectively. While larger EG droplets were found to be nearly hemispherical (e.g., see the large droplet to the left of the small drop of interest in Fig. 2(c)), smaller droplets tended to exhibit a smaller contact angle. From microscope images and the experimental thermomechanical spectrum, we estimated the droplet analysed in Fig. 2 to be a spherical 'cap' with base diameter ~ 126 μm and peak height ~ 47 μm.

Immediately obvious from inspection of the graph in Fig. 2(a) is that the noise spectrum is different in the two cases. Note that the 'bulk' spectrum (blue curve) is very typical for our 100 μm diameter sensors immersed in a liquid medium [11]. The 6 resonant features are the damped, circularly symmetric eigenmodes of the buckled membrane mirror. These eigenmodes have the highest overlap with the fundamental optical cavity mode used to interrogate the motion, and thus dominate the thermomechanical noise spectrum. Also note that the black curve in Fig. 2(a) is the photodetector shot noise recorded with the laser spatially misaligned from the cavity but for the same time-averaged received power. The separation between the blue and black curves indicates that, when the laser is aligned to the optomechanical sensor, the noise floor is almost entirely thermomechanical. In other words, the noise signal is then a faithful recording of the thermal Brownian motion of the membrane mirror. It is clear that the EG droplet spectrum (red curve) exhibits distinct differences from the bulk spectrum (blue curve), especially in the 20-40 MHz range where several 'extra' features are manifested. First, however, consider the general broadening and shifting (i.e., downwards in frequency) of the sensor vibrational modes when the amount of EG adjacent to the sensor is increased. This is as expected due to the additional 'added mass' and increased viscous damping for the 'bulk medium' case [11].

More interesting are the 'extra' spectral features above 20 MHz. To elucidate the origin of these features, we used a linear elastic fluids model and COMSOL software to numerically simulate the eigen-frequencies of the droplet viewed as an acoustic resonator (see the supplementary information file for details). Both the droplet-air and droplet-membrane interfaces were assumed to be 'soft' acoustic boundaries, while the portion of the droplet in contact with the substrate (i.e., the ring-shaped region outside the sensor) was assumed to be a 'hard' acoustic boundary. Values for the density and speed of sound in room-temperature EG were taken from the literature, and the dimensions of the EG droplet were estimated from microscope images. Notably, the use of a 'soft' boundary for the sensor interface was necessary to achieve a consistently good agreement between theory and experiment. This results in the acoustic modes having a pressure 'node' at the sensor; however, pressure nodes are locations of maximum particle (i.e., the EG fluid here) displacement in an acoustic resonator. Since our optomechanical sensor specifically detects displacement of the membrane mirror, this is a favorable situation. This also illustrates the intimate coupling between the droplet and membrane vibrational modes in this system – the membrane moves relatively freely in response to fluid motion which is in turn associated with the bulk compressive pressure modes of the droplet.

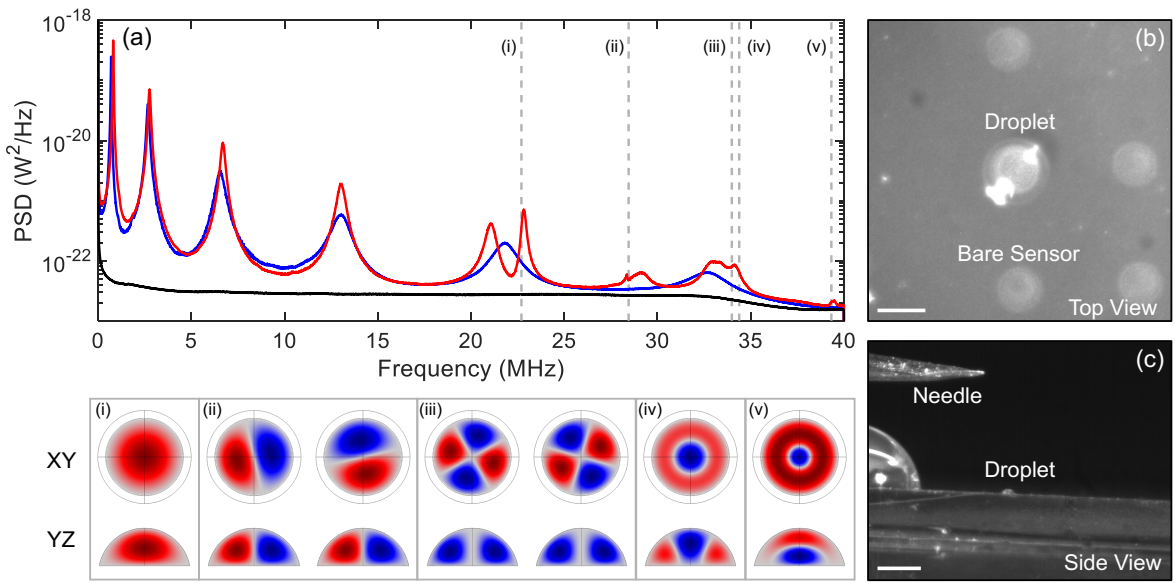

**Fig. 2.** (a) Power spectral densities for the same sensor in bulk EG (blue curve) and covered by a small EG droplet (red curve). Also shown is the shot noise floor for the same average received power at the photodetector (black), but in the absence of thermomechanical noise. The vertical dashed lines indicate the numerically predicted eigenfrequencies for the lowest-order bulk acoustic modes of the sessile droplet, and the images below the plot show the correspondingly predicted pressure distributions halfway up the droplet (XY) and an orthogonal plane (YZ) through the middle of the droplet. (b) Top-view microscope image of the droplet centered on the sensor of interest. Scale bar: 100 μm. (c) Side-view microscope image of the same droplet, with a much larger droplet visible at the left side of the image. Scale bar: 500 μm.

The 5 lowest-order eigenmode frequencies predicted by the simulation are overlaid as vertical dashed lines in Fig. 2(a). These align extremely well with resonant features visible in the experimental droplet spectrum, providing strong evidence that these features are in fact signatures of the droplet acoustic breathing modes. Slight discrepancies can be attributed mainly to the following: i. there is considerable error in estimating the droplet size/shape from the microscope images, ii. how well-centered a droplet is affects the boundary conditions in COMSOL, and iii. the COMSOL model treats the droplet as an independent resonator, whereas the experimental data exhibits clear evidence of hybridization between the mechanical modes of the droplet and the sensor [16]. Such hybridization produces Fano-like resonance features [20] which are not perfectly aligned with the independent oscillator modes of either of the resonators.

As a second example, analogous results for a larger EG droplet are shown in Fig. 3. In this case, base diameter ~ 516 μm and peak height ~ 258 μm was estimated. As above, COMSOL simulations predicted a fundamental resonance feature in very good agreement with the lowest-frequency 'extra' peak (in this case at ~ 4.5 MHz) observed in the thermomechanical noise spectrum of the sensor/droplet combination. The 15 lowest-order frequencies predicted by the simulation are shown by dashed lines in Fig. 3, and several (highlighted in black) overlap quite clearly with 'extra' features in the experimental spectrum. However, it becomes more challenging to confidently assign modes to features at higher frequencies. In addition to the two sources of discrepancy mentioned above, additional challenges arise for a larger droplet because it hosts a much denser collection of acoustic modes. Moreover, the predicted eigen-frequencies for the higher-order modes tend to be more sensitive to errors in the estimated droplet dimensions. Finally, as discussed elsewhere [16], the degree to which a given droplet mode appears in the thermomechanical noise spectrum of the sensor is dependent on a number of other details including the spatial and spectral alignment between that mode and the vibrational modes of the sensor itself. Nevertheless, from a large body of data collected across a range of droplet sizes (see the supplementary information file for additional examples), we have confidently established that the 'extra' features appearing in these noise spectra are in fact due to acoustic modes of the droplets interacting and hybridizing with the vibrational modes of the membrane element in our sensors.

It is worth emphasizing that the experimental observation of the acoustic breathing modes of a droplet has only recently been achieved [21], and in that case by actively exciting the droplet vibrations using a resonantly coupled laser. The data presented here provides remarkable evidence of the ability of an optomechanical sensor to operate at thermomechanical noise limits over a wide bandwidth, encompassing several of its own mechanical resonances [11-14], and even for devices with relatively modest optical and mechanical Q-factors (< $10^4$ for the sensors employed here). In this regime, the responsivity (e.g., the response to external pressure variations) and the noise floor are described by the same spectral dependence. For example, while the device is less responsive at off-resonance frequencies it also has a much lower noise floor at those same frequencies. The net result is that the noise-equivalent-pressure (NEP) of the sensor is relatively flat over the entire range for which thermomechanical noise is dominant [11,12] (i.e., up to ~40 MHz in our experiment, see Figs. 2 and 3). This enables detection of 'ambient-medium noise', dominated by resonant acoustic modes of the droplet over the entire range studied. Also remarkable, the displacement spectral density associated with these thermally driven modes is estimated to be less than 10-16 m/Hz$^{1/2}$ [16], and they nevertheless show up as high SNR Features in these plots.

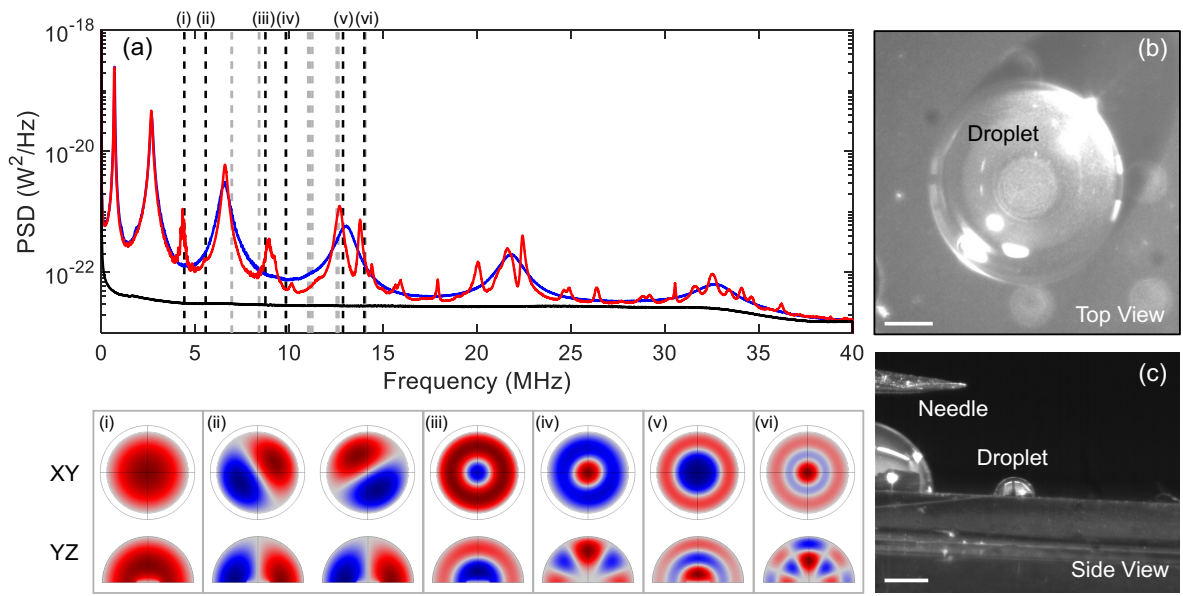

**Fig. 3.** Analogous results to those shown in Fig. 2, but for a larger droplet. (a) Experimentally measured noise spectra and several lowest-order, numerically predicted droplet eigenmodes. (b) Top-view microscope image. Scale bar: 100 μm. (c) Side-view microscope image. Scale bar: 500 μm.

For the boundary conditions described above, and for all droplet sizes studied, the fundamental droplet eigenmode is predicted to have the characteristics of a radial 'breathing' mode, with a pressure node (i.e., a displacement antinode) at both the droplet-air boundary and the droplet-sensor interface. This mode is therefore expected to be strongly coupled to the sensor, and this was borne out by the experimental data. Using dimensions estimated from microscope images, across a wide range of droplet sizes, we compared the numerically predicted fundamental droplet breathing mode frequency to the observed first 'extra' spectral feature in the thermomechanical noise spectrum. The results are shown in Fig. 4 and reveal an excellent agreement. The experimental data points include uncertainty bars, which correspond to our confidence in extracting the droplet dimensions from the microscope images (~ +/- 10 μm) and our confidence in extracting the resonant frequency from the noise plots (~ +/- 5 % of the frequency). Because of the variation in observed contact angle, the droplets vary in their shape, with larger droplets being very nearly hemispherical and smaller droplets better described as truncated spherical caps. Table S.1 in the supplementary information file gives the estimated radius and height for each droplet represented in Fig. 4, and those values were used in the simulations. In spite of the deviation from hemispherical shape for small droplets, the data in Fig. 4 could be well-fit to a simple scaling law [22]:

$$f_B \approx \frac{0.62 \cdot c}{h}. \qquad (1)$$

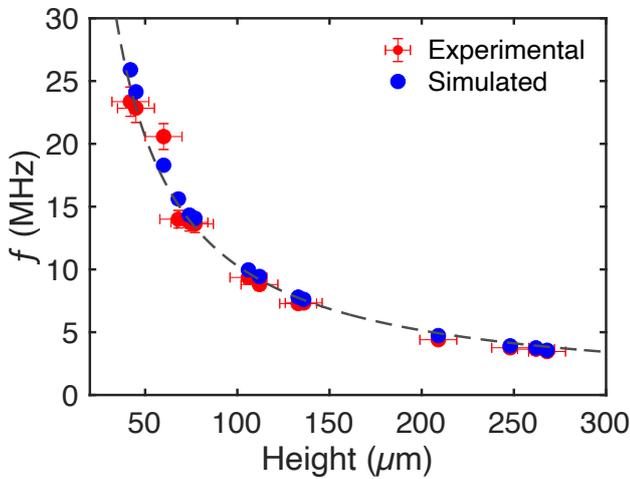

**Fig. 4.** Comparison of the experimentally observed (red) and numerically predicted (blue) resonance frequency for the lowest-order acoustic mode versus height of the droplet. The dashed line represents a simple scaling law fit as described in the main text.

Here, $c$ is the speed of sound in ethylene glycol ($c$ = 1660 m/s) and $h$ is the height of the droplet.

We also estimated the mechanical quality factors of the droplet acoustic modes from their resonant linewidths, and the results for the fundamental breathing mode are tabulated in the supplementary information file. As was the case for the extracted resonance frequencies, these estimates are approximate because the measured linewidths are impacted by the hybridization between droplet and sensor modes discussed above. Nevertheless, the estimated values ($Q$ ~ 20 – 100) are very similar to those previously reported for breathing modes of liquid droplets [16,21]. An analytical expression for Q-factor taking into account only the viscous damping within the droplet was provided by Dahan *et al.* [21], and predicts a Q-factor of several hundred for our EG droplets. However, it is well-known that several other dissipation mechanisms can contribute to damping of vibrational resonances, including radiation into surrounding media, thermal dissipation, and frictional effects [23]. For a sessile droplet, substrate interactions (and coupling to the optomechanical sensor itself) likely play a dominant role. A full study of these details is left for future work.

As a further test, we allowed EG droplets to slowly evaporate and captured both microscope images and the thermomechanical noise spectrum at a series of elapsed times. The physics of evaporating droplets is in its own right a topic of ongoing study [24] and technological importance, including as a means to induce mixing and particle aggregation in sessile-droplet-based systems [9-10]. A typical set of results is shown in Fig. 4, where a sensor was initially deposited with a slightly non-concentric EG droplet of diameter ~ 400 μm. As evidenced by the microscope images in Fig. 4(a), and not uncommonly [25], these droplets remained essentially pinned at a fixed base diameter over the entire monitored period (several hours) of evaporation. The estimated fundamental breathing mode frequency and volume of the droplet were recorded at intervals, with the results shown in Figs. 4(b) and (c), respectively. As evidenced by the data plotted in Fig. 4(d), the droplet shape and size are clearly correlated with resonant features in the thermomechanical noise spectra. As an aside, it can be observed that the resonant features in Fig. 4(d) are not as well-resolved as those in Figs. 2 and 3. We attribute this to the misalignment of the droplet and the sensor in this case, which leads to slightly reduced coupling between their vibrational modes. Techniques for precisely controlling the position, size, and shape of droplets are well known and would be a necessary precursor for most real-world applications of the proposed acoustic sensing approach.

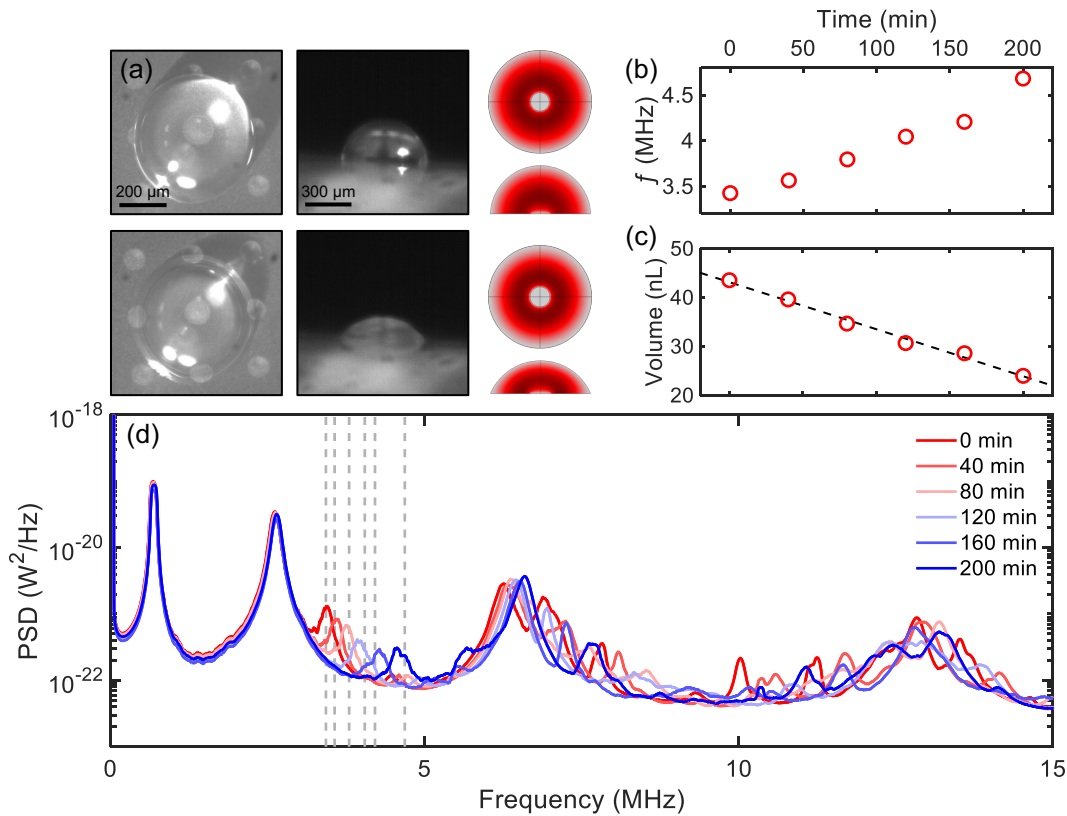

**Fig. 5.** (a) Top-view (left) and side-view (right) microscope images of a slightly misaligned droplet immediately after deposition (upper row) and after 200 minutes of evaporation (lower row). To the right of each set of images are top- and side-view depictions of the corresponding numerically simulated fundamental breathing mode. The XY plane cross-section was taken at the droplet-substrate interface. (b) Simulated frequency of the fundamental breathing mode versus evaporation time. (c) Droplet volume versus time as estimated from the microscope images. (d) The thermomechanical noise spectrum of the sensor is plotted for a series of measurement intervals as indicated in the legend. The vertical dashed lines indicated the numerically simulated fundamental breathing mode frequency for the corresponding droplet dimensions, estimated from microscope images captured at the same intervals.

## 4. Discussion and Conclusions

In summary, we have demonstrated the passive monitoring of thermally-driven acoustic (vibrational) modes of sessile droplets, by placing them in contact with on-chip optomechanical sensors. Since the speed and attenuation of sound are intimately correlated with the state and composition of a liquid, we believe this technique has potential to enable new approaches to chemical and biological assays in an open microfluidics platform. Moreover, many chemical and transformational phase changes are known to emit acoustic signals in the MHz frequency range [26], which suggests the potential for monitoring of dynamic processes within small-scale, droplet-based micro-reactors [27]. While this latter point is admittedly speculative, it appears to be fertile ground for exploration.

It is worth mentioning that passive ultrasonic spectroscopy is an established technique within larger-scale systems. In fact, it has been known for decades [28] that the noise signal in an ultrasound sensor contains information about its ambient environment [29]. This technique is used for example to extract information about the earth's crust from the noise in seismic sensors [30]. However, in conventional piezoelectric sensors and even in more recent capacitive sensors [31], ambient-medium noise is typically only a small fraction of the overall noise signal, which instead is dominated by electrical noise. This implies that extensive averaging and signal processing is required to extract information about the environment from the noise signal. Optomechanical sensors are uniquely enabling in this regard, because they can readily operate in a regime where ambient-medium noise makes a substantial contribution, in some cases even the largest contribution [13], to their noise floor. Moreover, they can achieve extreme sensitivity in a very small footprint, enabling them to act as passive probes within a micro-scale environment, such as the droplets studied here.

Finally, while the present results were restricted to detecting the vibrational motion of static sessile droplets, it is interesting to speculate whether optomechanical sensors might enable passive acoustic sensing of other micro- or nano-scale objects (perhaps encapsulated within a sessile droplet). Notably, solid (but relatively scarce) evidence exists that a wide range of biological entities such as bacteria [32], viruses, and microtubules [33] emit ultrasound waves (e.g., due to their natural modes of thermal vibrational motion or due to intracellular processes) at frequencies spanning from the KHz to GHz range. In spite of calls for more widespread study [34,35], limited experimental work has been reported in this area, presumably due to the rather extreme challenge of detecting these waves, which are inherently weak and often embedded in a background of acoustic noise arising from the environment. Nevertheless, Basiri-Esfahani *et al.* [13] have reported optomechanical sensors with similar sensitivity to those described here, and estimated that detecting the natural vibrational (acoustic) signatures of biological entities (bacteria, viruses, cells, DNA) was within reach. This type of 'acoustic spectroscopy' of small-scale objects could ultimately yield new insights into cellular and biological processes. Viewed globally, it seems clear that optomechanical sensors have potential to enable new strategies for research at the intersection between bio-acoustics and lab-on-a-chip systems.

## Author Contributions

RD, GH, and KS contributed to conception and design of the study. KS and FR performed the experiments and data analysis. RD wrote the first draft of the manuscript. KS wrote sections of the manuscript and the first draft of the supplementary information file. KS processed the data and generated the figures. KS and GH performed the simulations. All authors contributed to manuscript revision, have read, and approved the submitted version.

## Conflicts of interest

Ultracoustics Technologies Ltd. (I,P) KGS, North Road Photonics Corp. (E,P) GJH and (I,P) RGD.

## Acknowledgements

This research was funded by the Government of Alberta (Innovation Catalyst Grant), Alberta Innovates, the Natural Sciences and Engineering Research Council of Canada (CREATE 495446-17), and the Alberta EDT Major Innovation Fund (Quantum Technologies).

## References

1. R. C. Ng et al., "Excitation and detection of acoustic phonons in nanoscale systems," Nanoscale 14, 13428 (2022).
2. F. Eggers and U. Kaatze, "Broad-band ultrasonic measurement techniques for liquids," Meas. Sci. Technol. 7, 1-19 (1996).
3. U. Kaatze, F. Eggers, and K. Lautscham, "Ultrasonic velocity measurements in liquids with high resolution – techniques, selected applications and perspectives," Meas. Sci. Technol. 19, 062001 (2008).
4. A. P. Sarvazyan, "Developments of methods of precise ultrasonic measurements in small volumes of liquids," Ultrasonics 20(4), 151-154 (1982).
5. R. Fogel, J. Limson, and A. A. Seshia, "Acoustic biosensors," Essays in Biochem. 60, 101-110 (2016).
6. C. Peng, M. Chen, J. B. Spicer, and X. Jiang, "Acoustics at the nanoscale (nanoacoustics): a comprehensive literature review. Part I: materials, devices and selected applications," Sens. Actuators A 332, 112719 (2021).
7. J. Friend and L. Y. Yeo, "Microscale acoustofluidics: microfluidics driven via acoustics and ultrasonics," Rev. Mod. Phys. 83, 647-704 (2011).
8. W. Connacher et al., "Micro/nano acoustofluidics: materials, phenomena, design, devices, and applications," Lab Chip 18, 1952 (2018).
9. P. Dak et al., "Droplet-based biosensing for lab-on-a-chip, open microfluidics platforms," Biosensors 6, 020014 (2016).
10. J. L. Garcia-Cordero and Z. H. Fan, "Sessile droplets for chemical and biological assays," Lab Chip 17, 2150-2166 (2017).
11. G. J. Hornig, K. G. Scheuer, E. B. Dew, R. Zemp, and R. G. DeCorby, "Ultrasound sensing at thermomechanical limits with optomechanical buckled-dome microcavities," Opt. Express 30(18), 33083-33096 (2022).
12. B.-B. Li, L. Ou, Y. Lei, and Y.-C. Liu, "Cavity optomechanical sensing," Nanophotonics 10(11), 2799-2832 (2021).
13. S. Basiri-Esfahani, A. Armin, S. Forstner, and W. P. Bowen, "Precision ultrasound sensing on a chip," Nat. Commun. 10(1), 132 (2019).
14. W. J. Westerveld, Md. Mahud-Ul-Hasan, R. Shnaiderman, V. Ntziachristos, X. Rottenberg, S. Severi, and V. Rochus, "Sensitive, small, broadband, and scalable optomechanical ultrasound sensor in silicon photonics," Nature Photon. 15, 341-345 (2021).
15. A. M. Winkler, K. Maslov, and L. V. Wang, "Noise-equivalent sensitivity of photoacoustics," J. Biomed. Opt. 18(9), 097003 (2013).
16. G. J. Hornig, K. G. Scheuer, and R. G. DeCorby, "Observation of thermal acoustic modes of a droplet coupled to an optomechanical sensor", Arxiv
17. M. H. Bitarafan and R. G. DeCorby, "Small-mode-volume, channel-connected Fabry-Perot microcavities on a chip," Appl. Opt. 56, 9992-9997 (2017).
18. S. Al-Sumaidae, L. Bu, G. J. Hornig, M. H. Bitarafan, and R. G. DeCorby, "Pressure sensing with high-finesse monolithic buckled dome microcavities," Appl. Opt. 60(29), 9219-9224 (2021).
19. K. G. Scheuer and R. G. DeCorby, "All-Optical, Air-Coupled Ultrasonic Detection of Low-Pressure Gas Leaks and Observation of Jet Tones in the MHz Range", Sensors 23, no. 12: 5665 (2023).
20. A. E. Miroshnichenko, S. Flach, and Y. S. Kivshar, "Fano resonances in nanoscale structures," Rev. Mod. Phys. 82(3), 2257-2298 (2010).
21. R. Dahan, L. L. Martin, and T. Carmon, "Droplet optomechanics," Optica 3(2), 175-178 (2016).
22. J. L. Flanagan, "Acoustic modes of a hemispherical room," J. Acoust. Soc. Am. 37(4) 616-8 (1965).
23. V. Galstyan, O.-S. Pak, and H. A. Stone, "A note on the breathing mode of an elastic sphere in Newtonian and complex fluids," Phys. Fluids 27, 032001 (2015).
24. D. Zang, S. Tarafdar, Y.-Y. Tarasevich, M.-D. Choudhury, and T. Dutta, "Evaporation of a droplet: from physics to applications," Phys. Rep. 804, 1-56 (2019).
25. S. Sbarra, L. Waquier, S. Suffit, A. Lemaitre, and I. Favero, "Optomechanical measurement of single nanodroplet evaporation with millisecond time-resolution," Nature Comm. 13, 6462 (2022).
26. J. W. R. Boyd and J. Varley, "The uses of passive measurement of acoustic emissions from chemical engineering processes," Chem. Eng. Science 56, 1749-1767 (2001).

27 E. Salm, D. D. Guevara, P. Dak, B. R. Dorvel, B. Reddy, M. A. Aslam, and R. Bashir, "Ultralocalized thermal reactions in subnanoliter droplets-in-air," PNAS 110(9), 3310-3315 (2013).
28 R. L. Weaver and O. I. Lobkis, "Ultrasonics without a source: thermal fluctuation correlations at MHz frequencies," Phys. Rev. Lett. 87(13), 134301 (2001).
29 R. H. Mellen, "The thermal noise limit in the detection of underwater acoustic signals," J. Acoust. Soc. Am. 24(5), 478-480 (1952).
30 R. L. Weaver, "Information from seismic noise," Science 307, 1568-1569 (2005).
31 S. Lani, S. Satir, G. Gurun, K. G. Sabra, and F. L. Degertekin, "High frequency ultrasonic imaging using thermal mechanical noise recorded on capacitive micromachined transducer arrays," Appl. Phys. Lett. 99, 224103 (2011).
32 M. Matsuhashi et al., "Production of sound waves by bacterial cells and the response of bacterial cells to sound," J. Gen. Appl. Microbiol. 44, 49-55 (1998).
33 O. Kucera and D. Havelka, "Mechano-electrical vibrations of microtubules - link to subcellular morphology," BioSystems 109, 346-355 (2012).
34 G. Reguera, "When microbial conversations get physical," Trends in Microbiol. 19, 105-113 (2011).
35 S. Barzanjeh, V. Salari, J. A. Tuszynski, M. Cifra, and C. Simon, "Optomechanical proposal for monitoring microtubule mechanical vibrations," Phys. Rev. E 96, 012404 (2017).